\documentclass[aps,prl,twocolumn,groupedaddress,showpacs]{revtex4}
\usepackage{graphicx}
\usepackage{amsmath}
\usepackage{natbib}
\usepackage[usenames]{color}
\begin{document}
\title{The Poincar\'e--Birkhoff theorem in Quantum Mechanics}
\author{D. A. Wisniacki}
\affiliation{ Departamento de F\'\i sica and IFIBA,
FCEyN, UBA Ciudad Universitaria,
Pabell\'on 1, Ciudad Universitaria, 1428 Buenos Aires, Argentina.}

\author{M. Saraceno}
\affiliation{Departamento de F\'{\i}sica,
 Comisi\'on Nacional de Energ\'{\i}a At\'omica.
 Av.~del Libertador 8250, 1429 Buenos Aires, Argentina.}

\author{F. J. Arranz}
\affiliation{Grupo de Sistemas Complejos
 and Departamento de F\'{\i}sica,
 Escuela T\'ecnica Superior de Ingenieros
 Agr\'onomos, Universidad Polit\'ecnica de Madrid,
 28040 Madrid, Spain.}

\author{R. M. Benito}
\affiliation{Grupo de Sistemas Complejos
 and Departamento de F\'{\i}sica,
 Escuela T\'ecnica Superior de Ingenieros
 Agr\'onomos, Universidad Polit\'ecnica de Madrid,
 28040 Madrid, Spain.}

\author{F. Borondo}
\affiliation{Departamento de Qu\'{\i}mica, and
 Instituto Mixto de Ciencias Matem\'aticas CSIC-UAM-UC3M-UCM,
 Universidad Aut\'onoma de Madrid,
 Cantoblanco, 28049--Madrid, Spain.}
\date{\today}
\begin{abstract}
Quantum manifestations of the dynamics around resonant tori in
perturbed Hamiltonian systems, dictated by the Poincar\'e--Birkhoff theorem,
are shown to exist.
They are embedded in the interactions involving states which differ
in a number of quanta equal to the order of the classical resonance.
Moreover, the associated classical phase space structures are
mimicked in the quasiprobability density functions and their zeros.
\end{abstract}
\maketitle
%
The quantum manifestations of classical chaos,
or \emph{quantum chaos}, have received much attention in the
recent past \cite{qchaos}.
Besides random matrix theory, which describes universal properties,
Gutzwiller's trace formula describes the dual relationship between
coherent sums over periodic orbits with sums over quantized states \cite{chaosbook}.
The appearence of scars of periodic orbits in individual eigenfunctions
and the associated scar theory are results that deserve special mention
\cite{Houches,Heller,homo}. These results are by now routinely applied
in a variety of important technological applications \cite{nano}.
At the other extreme, for integrable systems, this dual relationship
is more precise and collective sums over the degenerate periodic orbits
on rational tori give rise to individual states on quantized tori \cite{berry}.

For systems close to integrable, the classical behavior is
completely understood in terms of the celebrated
Kolmogorov-Arnold-Moser (KAM) and the Poincar\'e-Birkhoff (PB)
theorems \cite{KAMclasico,PB}.
The first one deals with quasiperiodic motion and the persistence
of sufficiently irrational tori under perturbations.
The second considers the fate of the (unperturbed) resonant tori,
from which an even number of POs survive.
In their vicinity, chains of islands of regularity surrounded by
a chaotic separatrix organized by a homoclinic tangle are formed.
This structure and the associated transport are well described by
the universal pendulum model of Chirikov \cite{Chirikov},
being the (slow) Arnold diffusion \cite{KAMclasico}
the controlling mechanism of transport.
Classical resonances are known to control many relevant processes,
such as intramolecular vibrational relaxation \cite{Wolynes1},
or directional laser emission in nanooptics \cite{nano,nanoptics}.
On the other hand, the analysis of the quantum counterpart has
produced much fewer results \cite{Taylor,Geisel,Wolynes2}.

In this Letter, we explore the quantum implications of the
classical PB theorem.
We find that they exist and can be unveiled by studying the
mechanism by which two quantized tori of the unperturbed system interact
to form the beginning of an island chain on a resonant rational torus.
The subtle mechanism by which the quantum numbers of the unperturbed tori are
exchanged in a Landau-Zener (LZ) \cite{LandauZ} quasi-crossing is illustrated
by the exchange of zeroes of the stellar representation \cite{voros}.
The splittings of a series of avoided crossings (ACs) ruled by a PB resonance
in the correlation diagram is explained semiclassically.

The model that we have chosen to study is the Harper map in the unit square,
%
\begin{eqnarray}
  p_{n+1} & = & p_n + k \sin(2\pi q_n) \qquad ({\rm mod}\; 1), \nonumber \\
  q_{n+1} & = & q_n - k \sin(2\pi p_{n+1}) \qquad ({\rm mod}\; 1),
  \label{eq:1}
\end{eqnarray}
where $k$ is a parameter measuring the strength of the perturbation.
The quantized version is provided by the unitary time-evolution operator
\cite{Leboeuf1,Paz}
%
\begin{equation}
  \hat{U} = \exp[{\rm i}N k \cos(2\pi \hat{q})] \; \exp[{\rm i}N k \cos(2\pi \hat{p})],
  \label{eq:3}
\end{equation}
with $N=(2\pi\hbar)^{-1}$.
The simplicity of this model allows extremely detailed calculations.
\begin{figure}
 \includegraphics[width=7cm]{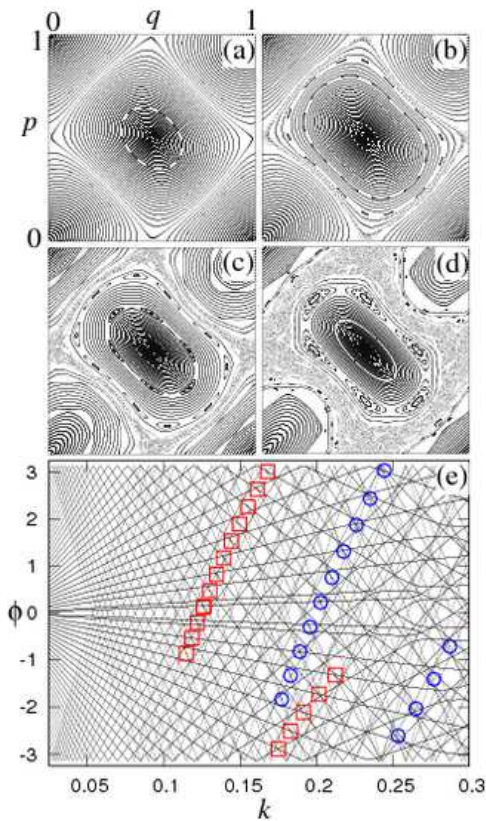}
 \caption{(color online)
   Classical phase space for the Harper map (\ref{eq:1})
   for $k$: (a) 0.1, (b) 0.155, (c) 0.2, and (d) 0.25. \\
   (e) Correlation diagram for the eigenphases of the quantized Harper map
   (\ref{eq:3}) with $N=60$ as a function of the perturbation parameter $k$.
   Two series of avoided crossings of states with quantum numbers differing in
   10 (red squares) and 6 (blue circles) units, respectively, are shown.}
 \label{fig:1}
\end{figure}
The classical dynamics of this map follows the KAM and PB
pattern as a function of $k$,
depicted in Fig.~\ref{fig:1}(a)-(d). For small values $k=0.1$ [panel (a)]
all trajectories look regular at the scale
of the figure, being confined to invariant tori.
For $k=0.155$ [panel (b)] two resonances of order 1:14 and 1:10
around the center become important and the corresponding chains of
islands are visible.
Also, the separatrix between central tori and those at corners shows as an
appreciable band of stochasticity.
As $k$ reaches 0.2 [panel (c)] the classical phase space continues evolving;
the 1:14 resonance disappear, swallowed by chaos in the main separatrix,
the 1:10 resonance grows in size, and a new 1:6 resonance becomes apparent.
Finally, for $k=0.25$ [panel (d)] only the 1:6 resonance survives.

The quantum map can be diagonalized resulting in a set of $N$ eigenphases,
$\phi_n \in [-\pi, \pi)$, and eigenfunctions, $|\phi_n\rangle$.
Plotted as a function of $k$ the eigenphases produce the correlation diagram
in Fig.~\ref{fig:1}(e).
To interpret this diagram we rewrite the map using the Baker-Hausdorff
expansion for non-commuting exponentials \cite{bakerhaus,scharf} as
%
\begin{equation}
  \hat{U} = \exp[{\rm i}N k \hat H^{(n)} (k)] \; \exp[{\rm i}N V^{(n)}(k) ],
  \label{eq:4}
\end{equation}
where $\hat H^{(n)}=\hat H_0+k \hat H_1 + \cdots k^{n-1}\hat H_{n-1}$
and $\hat V^{(n)}(k)$ is another operator with a power series expansion starting
with the power $k^{n+1}$.
All terms of both expansions can be computed in terms of iterated commutators of
$\cos(2\pi \hat{q})$ and $\cos(2\pi \hat{p})$.
If commutators are replaced by Poisson brackets in the usual fashion,
the equivalent classical expansion is obtained \cite{scharf}.
As an example the classical expansion to second order yields
%
\begin{equation}
\hat H^{(2)}(p,q)=\frac{\cos(2\pi \hat q)+\cos(2\pi \hat p)}{2\pi}
  +\frac{k}{2}\sin(2\pi \hat q)\sin(2\pi \hat p)
\end{equation}
\begin{equation*}
\hat V^{(2)}(p,q)=-\frac{k^2}{12}\left[\cos(2\pi \hat q)\sin^2(2\pi \hat p)+
\cos(2\pi \hat p)\sin^2(2\pi \hat q)\right]
\end{equation*}
Further terms can be (laboriously) calculated.
The importance of this rewriting for our purposes is that for moderate values of $k$
the  exact eigenphases can be interpreted as the levels of the integrable operator
$ \hat H^n(k) $ with definite quantum numbers $n$.
When multiplied by $kN$ and wrapped around the unit circle, the wrapping produces ACs
of states with different quantum number which become ACs of LZ type
\cite{LandauZ} when the small terms of order $k^{n+1}$ in $\hat V$ are taken into account.
This procedure becomes eventually divergent reflecting the fact that the chaotic behavior
of the map cannot be captured by any integrable approximation. However the procedure is well
suited to describe the beginning of the PB process when a rational torus disintegrates
in an island chain.
This situation is clearly visible in the correlation diagram of the eigenphases,
calculated for $N=60$.
There are 60 eigenlines all starting at zero for $k=0$,
and then they spread in an fan-like fashion, first linearly with $k$.
The slopes are given initially by the eigenvalues of
$\hat H_0=[\cos(2\pi \hat{q}) +\cos(2\pi \hat{p})]/2\pi$ and they are ordered in
increasing value of their quantum number.
Half of these lines have positive slope corresponding to states localized around
$(p,q)=(0,0)$ while the ones with negative slopes are centered on $(p,q)=(1/2,1/2)$.
We will restrict our analysis to the island formation in this latter case,
the other one being equivalent since connected by a simple symmetry operation.
%
%
%
\begin{figure*}
 \includegraphics[width=16cm]{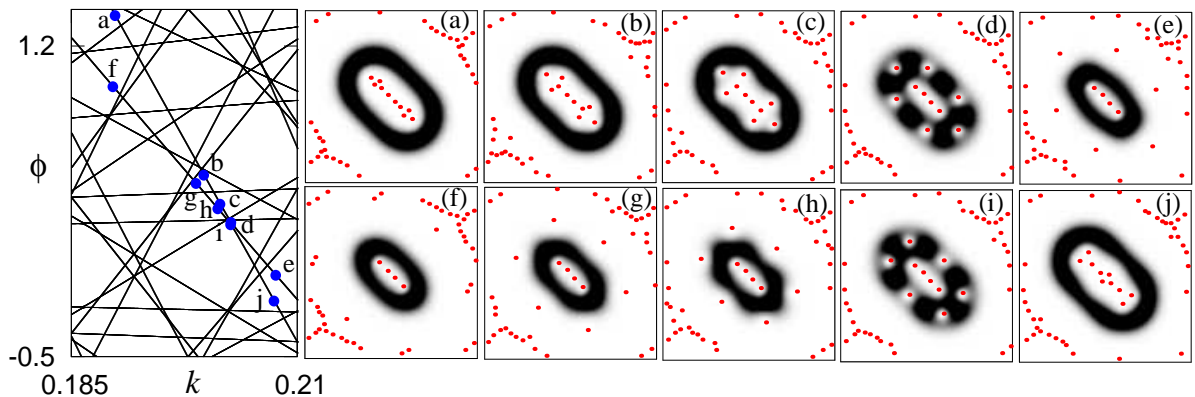}
 \caption{(Left) Detail of the correlation diagram of Fig.~\ref{fig:1}
   corresponding to the fifth element of the family of avoided crossings
   with $\Delta n=6$ which involves states with quantum numbers
   $n_1=4$ and $n_2=10$. \\
   (Right) Evolution of the corresponding quasiprobability densities
   given by the Husimi functions at selected points [(a)-(j)]
   in the diagram.
   Full circles mark the position of the zeros of the Husimi functions.}
 \label{fig:2}
\end{figure*}

Two interesting features arise from the analysis of the correlation diagram.
First, the ACs always appear grouped in series or families,
where the involved states differ by the same number of quanta, $\Delta n$.
Two such families of ACs, corresponding to $\Delta n=10$ and $\Delta n=6$
have been identified and marked with red squares and blue circles,
respectively, in the correlation diagram of Fig.~\ref{fig:1}.
The detailed analysis of one of these AC is shown in Fig.~\ref{fig:2}.
The AC involves states with $n_{\rm lower}=4$ and $n_{\rm upper}=10$
in the $\Delta n=6$ family and is a typical sharp two-level
LZ AC \cite{LandauZ}.
The left panel of Fig.~\ref{fig:2} is a blow-up of Fig.~\ref{fig:1}
in the range $k=0.185-0.21$.
Points (a)-(f), (b)-(g), (c)-(h)(d)-(i) and (e) and (j) show different stages
of the AC, the minimum $\Delta \phi$ being attained at point (d)-(i).

More interesting is the evolution of the associated Husimi functions,
which is shown, using the same labelling, in the right panels of the figure.
Recall that the Husimi distribution of a quantized torus is localized on it,
and its quantum number is given by the number of zeroes enclosed by the torus
\cite{Leboeuf1,voros}.
The rest of the zeroes, in this case to a total of $N=60$,
is scattered in the complementary phase space region.
Far from the AC, at points [(a)-(f), (b)-(g)] the two tori maintain their
individuality, but the zeroes start moving in a very characteristic way:
the lower state $n=4$ attracts $6$ zeroes while $6$ zeroes start migrating
outwards in the upper $n=10$ state.
The process culminates in panels (d)-(i) where the two states are now precisely
localized on the rational torus but with maxima either on the stable or the
unstable periodic points given by the PB theorem (the upper state sits on top
of the unstable points and the lower on the stable points).
Similarly, the position of the zeroes alternate between the fixed points,
localizing in the immediate neighborhoods.
Beyond the AC, the original topology is restored,
aside for the compulsory exchange of quantum numbers.
We have confirmed this mechanism for all other ACs of this family and also
for the 1:10 resonance.
We can then conclude that this is a general mechanism by which two quantized
tori interact via the periodic orbits - stable and unstable - that,
according to the PB theorem, remain on the resonant torus.
Even more interesting is the fact that the number of zeroes exchanged
between the tori is precisely the order of the resonance.

\begin{figure}
 \includegraphics[width=7cm]{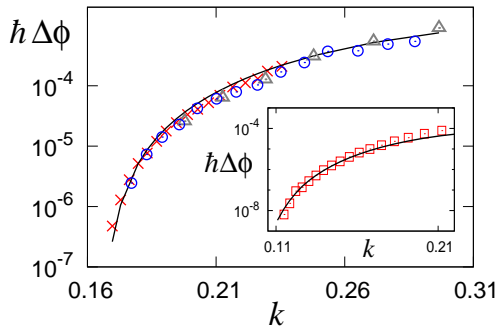}
 \caption{(color online)
   Numerical eigenphase differences, $\Delta \phi$, for avoided crossings
   of the $\Delta n=6$ family extracted from the correlation diagram of
   Fig.~\ref{fig:1}(blue circles) as a function of the perturbation
   parameter for $N$:
   (gray $\bigtriangleup$) 30, (blue $\odot$) 60, (red $\times$) 100,
   and (full line) semiclassical calculations based on Eq.~(\ref{eq:4})
   for the 1:6 resonance. \\
   Inset: Same for the 1:10 resonance and $N=100$.}
 \label{fig:3}
\end{figure}

Let us now make the above arguments quantitative.
%
%
%
At a LZ AC the splitting of the levels is given by the interaction
matrix element between the two involved states.
Quantum mechanically it could be calculated from the eigenstates of the integrable
hamiltonian $H^{(n)}$ perturbed by $V^{(n)}$ in Eq.~(\ref{eq:4}) and computed up
to a given power of $k$.
However, we are interested in a purely semiclassical calculation.
To avoid the very cumbersome transformation to action-angle variables,
in which the calculation would be trivial, we resort to a method
developed in Ref.~\cite{RAT} in connection with the calculation of
tunneling rates in this same model.
%
%
The resulting expression for the eigenphase difference at the AC due to a
$1:r$ resonance is then obtained in terms of purely classical information as
%
\begin{equation}
  \hbar \Delta \phi = \frac{1}{8 \pi^4 r}
     \int_0^{2\pi} \exp(-i r \theta) \; \delta I_{1:r}(\theta) d\theta,
  \label{eq:5}
\end{equation}
$\delta I_{1:r}(\theta)$ being given by
%
\begin{equation}
   \delta I_{1:r}(\theta) = I^{(-1)}(I_{1:r},\theta)-I_{1:r},
  \label{eq:6}
\end{equation}
where $I^{(-1)}(I_{1:r},\theta)$ symbolizes  the action variable that
is obtained by applying the inverse Poincar\'e map to $(I,\theta)$ or,
alternatively, the backward propagation with $H$ from time
$t=\tau$ to $t=0$.
The way the calculations are done in practice is the following.
First, the resonant periodic torus is located, and its action computed.
This is done by using an analytical third order approximation to the
Harper Hamiltonian in Eq.~(\ref{eq:4}), followed by numerical refinement
propagating trajectories near the torus.
Once the resonant torus is found, the corresponding angle variable is
computed for a large sample of points, which are then
back-propagated with the exact inverse map.
(This is a trivial process since it just implies exchanging the
$p$ and $q$ variables.)
The associated action, corresponding to each of these points, is then
calculated by numerical propagation of the trajectories in a complete cycle.
This provides the values of the perturbed action,
from which $\Delta \phi$ is evaluated with the aid of Eq.~(\ref{eq:5}).
The corresponding results for the 1:6 resonance are shown
as a continuous line in Fig.~\ref{fig:3}.
As can be seen, the agreement of this semiclassical calculation with
the data (triangles) extracted from the correlation diagram of
Fig.~\ref{fig:1} is extremely good, this giving full support to our
previous qualitative arguments on the existence of a clear quantum
manifestation of the classical PB theorem.
%

A final point worth addressing is the limit $N \rightarrow \infty$
 of our results,
which is closely related with the extent to which quantum mechanics
can distinguish the PB structure of a given classical resonance.
According to KAM and PB theorems such structures exist even
for infinitesimal values of the perturbation,
although as $k$ decreases they span smaller regions of phase space.
On the other hand, the finiteness of $\hbar$ imposes a limit to the
phase space details which can be resolved quantum mechanically.
How do they two extremes reconcile with our previous conclusions
is not clear in principle.
For example, in the correlation diagram of Fig.~\ref{fig:1}
the 1:6 resonance is only observed for $k > 0.166$, point at which the
first element, i.e.~$n_{\rm lower}=0$, of the corresponding family
of ACs is located.
Could this resonance be resolved for smaller perturbations if a
smaller value of $\hbar$ was chosen?
Fortunately, this question can be easily answered in our calculations,
where $\hbar$ is a parameter directly related to the size of the
Hilbert space, $N=(2\pi \hbar)^{-1}$.
Indeed, the answer to the question is yes, as demonstrated by Fig.~\ref{fig:3},
where the results for $\Delta \phi$ for three different values of $N$
are shown.
For increasingly larger values of $N$ the first member
of the ACs series appears at smaller values of $k$,
 the interaction being satisfactorily described in all cases by the
semiclassical theory.

Finally, let us remark that our conclusions are also valid for other
resonances and systems.
This is illustrated in the inset to Fig.~\ref{fig:3}, where the results
for the resonance 1:10 are shown, and a perfect agreement
between quantum and semiclassical results is also seen.
Moreover, a similar study has been carried out in a realistic model for the
vibrational dynamics of the LiCN molecule \cite{LiCN} arriving at the same
conclusions; the results will be reported elsewhere.

Summarizing, in this Letter we have shown that quantum
manifestations of the PB theorem,
a cornerstone to rationalize classical Hamiltonian chaos,
exist.
They can be unveiled by analyzing the interactions,
or ACs splitting in a correlation diagram,
involving pairs of states which differ in a number of quanta equal
to the order of a given classical resonance.
This result has remarkable effects,
since as a (perturbational) parameter varies, and a quantized
torus passes through the resonance it looses (or gains) zeroes
in its Husimi representation.
At resonance the zeroes are located at the position
of the periodic points predicted by the PB theorem, and the number of
zeroes, lost or gained, is equal to the order of the resonance.
Our conclusions are supported with semiclassical calculations
of the relevant ACs splitting ruled by the PB resonance.

\begin{acknowledgments}
This work was supported by the MICINN-Spain under contracts
\emph{iMath}--CONSOLIDER CDS2006--32 and MTM2009--14621, UBACyT (X237)
and CONICET (Argentina).
\end{acknowledgments}
%

%
\end{document}